\documentclass[preprint,12pt]{elsarticle}

\usepackage{amssymb}
\usepackage[utf8]{inputenc}
\usepackage[english]{babel}
\usepackage{amsmath}
\usepackage{amsfonts}
\usepackage{makeidx}
\usepackage{graphicx}

\usepackage{graphicx}
\usepackage{longtable}
\usepackage{float}
\usepackage{dcolumn}
\usepackage{bm}
\usepackage{appendix}
\usepackage{multirow}

\journal{Physics Letters B}

\begin{document}

\begin{frontmatter}



\title{Observational constraints on $\alpha$-attractor inflationary models with a Higgs-like potential}


\author[a]{J. G. Rodrigues} \ead{jamersoncg@gmail.com}

\author[b]{S. Santos da Costa}

\author[b]{J. S. Alcaniz}

\address[a]{Departamento de F\'isica, Universidade Federal do Rio Grande do Norte, 59072-970, Natal, RN, Brazil}

\address[b]{Observat\'orio Nacional, 20921-400, 
Rio de Janeiro, RJ, Brazil}

\begin{abstract}

We investigate the observational viability of a class of $\alpha$-attractors inflationary models in light of the most recent Cosmic Microwave Background (CMB) and Large-Scale Structure (LSS) data. By considering a double-well potential we perform a slow-roll analysis to study the behavior of this class of models, which is a continuous interpolation between the chaotic inflation for large values of $\alpha$ and the universal attractor, i.e., $n_s=1- 2/N$ and $r=\alpha 12/N^2$ for small $\alpha$, where $n_s$ is the scalar spectral index, $r$ is the tensor-to-scalar ratio, and $N$ is the e-fold number. In order to explore the parameter space of the model, we also perform a MCMC analysis and find $\alpha=7.56\pm 5.15$ ($1\sigma$).

\end{abstract}



\begin{keyword}

Inflation \sep Cosmological Parameters from CMB \sep MCMC Analysis



\end{keyword}

\end{frontmatter}


\section{Introduction}
\label{Introduction}

The inflationary paradigm has become a central part of modern cosmology as it provides a successful description of the first instants of the evolution of our  Universe~\cite{Guth:1980zm,Linde:1981mu,Albrecht:1982wi}. Since it was proposed, a considerable number of inflationary models were considered and investigated in light of observational data~\cite{Martin:2013tda}. From the current observational perspective, the most promising scenarios rely on modified gravity theories, such as the Starobinsky model $R+R^2$ \cite{Starobinsky:1980te,mukhanov1981quantum,Whitt:1984pd}, where $R$ is the Ricci scalar, the chaotic inflation $\lambda \phi^4$ non-minimally coupled to gravity  \cite{Salopek:1988qh,Fakir:1990eg,Makino:1991sg} and a class of super-conformal inflationary theories that leads to a universal attractor behaviour \cite{Kallosh:2013hoa,Kallosh:2013tua,Kallosh:2013yoa}.

Indeed, the most recent Cosmic Microwave Background (CMB) and Large Scale Structure (LSS) data favour models which present a plateau in their inflationary potentials~\citep{Akrami:2018odb,Martin:2013nzq}. In particular, the Starobinsky and the non-minimal models exhibit a similar potential in the regime of large fields and strong couplings~\citep{Linde_2011,Kehagias:2013mya},
\begin{equation} 
V(\phi) = \frac{3}{4} M^2 \left(1-e^{-\sqrt{\frac{2}{3}}\phi} \right)^2, \label{StarLike}
\end{equation}
where the inflaton $\phi$ is given in Planck units, $M_P= 1$. The energy scale $M$ is usually constrained by the observed value of the amplitude of scalar perturbations $A_S \simeq 2.1 \times 10^{-9}$ \cite{Aghanim:2018eyx}. For large values of the e-fold number $N \gg 1$, i.e. the number of $e$-foldings between the horizon crossing of modes of interest and the end of inflation, these theories lead to the same predictions for the inflationary observables, 
\begin{equation}
    n_s = 1- \frac{2}{N}, \quad r = \frac{12}{N^2},\label{StarLikeObservable}
\end{equation}
where $n_s$ is the scalar spectral index and $r$ the tensor-to-scalar ratio. For $50 \leq N \leq 60$, the expressions above yields $0.96 \lesssim n_s \lesssim 0.97$ and $0.003 \lesssim r \lesssim 0.005$, showing a remarkable agreement with present CMB observations~\cite{Akrami:2018odb,Aghanim:2018eyx}.

Given their ability to describe the observed data, these inflationary scenarios and their phenomenological implications have been extensively studied \cite{Kehagias:2013mya,Bezrukov:2007ep,Barvinsky:2008ia,GarciaBellido:2008ab,Bezrukov:2014bra,Hamada:2014iga,Gomes:2016cwj,Lee:2018esk}. In particular, non-minimal inflation  with the Higgs field playing the role of inflaton have received a fair amount of attention, as it can potentially connect inflationary dynamics with the low-energy phenomenology of the model.  Scenarios in which extensions of the standard model drive the inflationary dynamics have also been considered in the literature \cite{Lerner:2009xg,Okada:2011en,Ballesteros:2016euj,Ferreira:2017ynu}.

More recently, a class of inflationary models introduced by Kallosh, Linde and Roest has received considerable attention 
\cite{Kallosh:2013yoa,Kallosh:2014rga,Kallosh:2015lwa,Roest:2015qya,Linde:2016uec,Terada:2016nqg,Ueno:2016dim,Odintsov:2016vzz,Akrami:2017cir,Dimopoulos:2017zvq,Pozdeeva:2020shl,Odintsov:2020thl}. Based on supergravity, the so called cosmological $\alpha$-attractors stand out for achieving a slow-roll phase, even for inflationary potentials of arbitrary form $V(\phi)$ \cite{Linde:2016uec}. Specifically, an inflationary plateau arise in the theory due to the presence of a non-canonical kinetic term for the inflaton $\phi$. Such kinetic energy features a pole at $\phi^2 = 6\alpha$, which prevents any trajectory in field space to travel beyond this limit. On the other hand, a canonical form for the Lagrangian density can be achieved after a field redefinition. In the canonical field space the poles are displaced to infinity, ``stretching" the scalar potential in the large field regime. Similarly to the Starobinsky inflation, such plateau at high energies aligns the model predictions with the observational data.


Despite of all the promising characteristics of the $\alpha$-attractor class of models, only a few works address a careful statistical analysis of this scenario (see \cite{Akrami:2017cir,Garcia-Garcia:2018hlc,Cedeno:2019cgr} for the analysis of quintessential $\alpha$-attractor models). In this sense, we aim to deepen the analysis of these models, investigating its observational viability in light of the most recent CMB and LSS data ~\citep{collaboration2019planck,bao1,bao2,bao3,bicep21,bicep22}. To this end, we use the double-well function as non-canonical potential $V(\phi)$. Such a choice finds its motivation in fundamental particle physics, where this potential can be employed in the spontaneous symmetry breaking mechanism \cite{Goldstone:1961eq,Higgs:1964ia,Higgs:1966ev}.

This work is organized as follows. Section \eqref{model} reviews the basics of $\alpha-$attractors, discussing the field redefinition to obtain the canonical inflationary Lagrangian density. Section \eqref{slowrollanalysis} presents the slow-roll analysis for the double-well potential. In Section \eqref{analysisandresults} we present the observational data sets used in our statistical analysis and the main results obtained. Finally, the main conclusions of the paper are summarized in Section \eqref{conclusions}. 

\section{The Model}\label{model}

The Lagrangian density for the $\alpha$-attractor class of models arise in the context of supergravity theories for specific forms of the K\"ahler potential,
\begin{equation}
    \mathcal{L}= \sqrt{-g}\left[\frac{1}{2}R - \frac{1}{(1-\phi^2/6\alpha)^2}\frac{(\partial \phi)^2}{2}- V(\phi) \right], \label{attractorlagran}
\end{equation}
where $\phi$ is the inflaton field and $\alpha$ the free parameter of the theory, related to the curvature of the K\"{a}hler manifold \cite{Kallosh:2013yoa}. The main characteristics of these models manifests due to the presence of a non-canonical kinetic term for the inflaton. It exhibits a pole at $\phi^2 = 6\alpha$, preventing any displacement in the field space from crossing this limit. On the other hand, a canonical kinetic term can be recovered through the field redefinition,
\begin{equation}
    \phi = \sqrt{6\alpha}\tanh{\frac{\varphi}{\sqrt{6\alpha}}}, \label{fieldrelation}
\end{equation}
according to which the inflationary Lagrangian density assumes the form,
\begin{equation}
    \mathcal{L}= \sqrt{-g}\left[\frac{1}{2}R - \frac{(\partial \varphi)^2}{2}- V(\sqrt{6\alpha}\tanh{\frac{\varphi}{\sqrt{6\alpha}}}) \right]. \label{attractorlagran2}
\end{equation}

While the field $\phi$ is limited by the poles $\phi = \pm \sqrt{6\alpha}$, the canonical inflaton $\varphi$ is free to assume any value in field space. As result, the scalar potential $V(\sqrt{6\alpha}\tanh{\frac{\varphi}{\sqrt{6\alpha}}})$ acquire  a plateau in the limit of large fields, $\varphi \rightarrow \pm \infty$. 
Such configuration is propitious to provide large field inflation, leading to the following predictions for the inflationary observables,
\begin{equation}
    n_s = 1- \frac{2}{N}, \quad \text{and} \quad r = \alpha \frac{12}{N^2},\label{AttractorObservable}
\end{equation}
for large $N$ and small $\alpha$.

Similarly to the Starobinsky model, the $\alpha$-attractors theoretical predictions are aligned with observations, at least for the slow-roll analysis. Indeed, the predictions of both scenarios coincide at leading order for $\alpha = 1$. This predictive pattern is not a coincidence, as the inflationary potential in both cases is similar at the limit of large field.  
Particularly, in the vicinity of $\phi=\sqrt{6\alpha}$, or equivalently $\varphi \rightarrow \infty$, the potential for the canonical scalar field can be written in the asymptotic form
\begin{equation}
    V(\varphi) \simeq V_0-2\sqrt{6\alpha}V^\prime_0\exp{(-\sqrt{\frac{2}{3\alpha}}\varphi)}. \label{attractorlagran3}
\end{equation}
The expression above is reduced to the Starobinsky potential (\ref{StarLike}) for $\alpha = 1$ at leading order, except for an unobservable phase shift in $\varphi$. In this sense, one might identify the Starobinsky potential as one of the solutions of the $\alpha$-attractor class, reinforcing the general aspect of these inflationary models.

%
\begin{figure}[!ht]
 \begin{center}
 \includegraphics[scale=0.6]{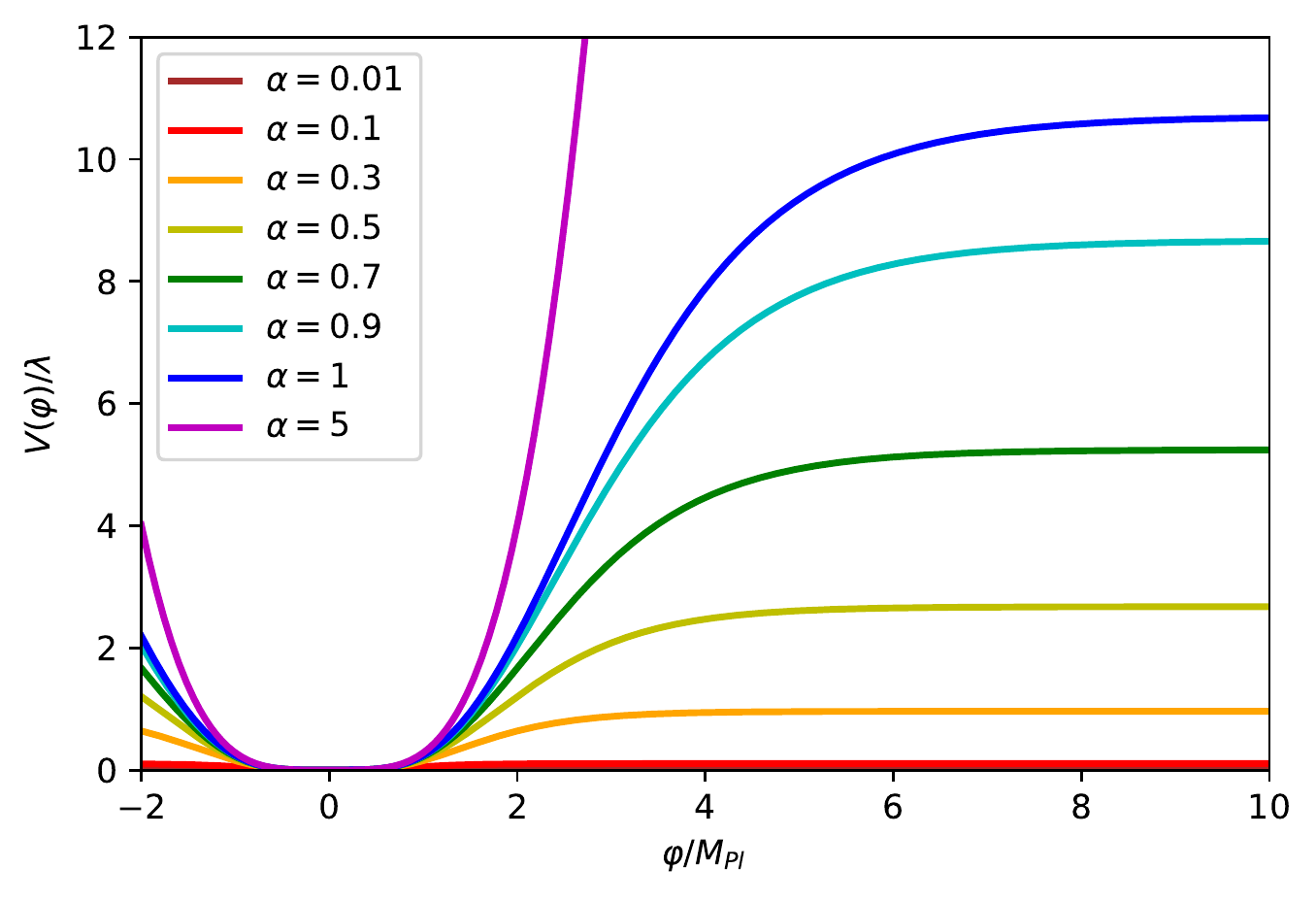}
 \end{center}
\caption{The Higgs-like primordial potential as a function of the canonical field $\varphi$ for different values of $\alpha$.
}
\label{fig:higgs_sta_pot}
\end{figure}

Despite the general aspect of the $\alpha$-attractor models at small values of $\alpha$, it is important to consider a specific choice of inflationary potential in order to correctly account for the predictions of the model for a broad range of $\alpha$. Therefore, the slow-roll and statistical analysis of this work will focus on the double-well inflationary potential,
\begin{equation}
    V(\phi) = \frac{\lambda}{4}(\phi^2-v^2)^2, \label{doublewell}
\end{equation}
where $\phi$ is a real inflaton and $v$ its vacuum expectation value. 

This choice of primordial potential is strongly motivated in particle physics, where it can be applied in the dynamical symmetry breaking mechanism \cite{Goldstone:1961eq,Higgs:1964ia,Higgs:1966ev}. In particular, the scalar sector of the standard model of fundamental particles is composed of such potential, with the scalar field behaving like a doublet under the $SU(2)_L$ symmetry of the standard model, namely the Higgs doublet \cite{Weinberg:1967tq,Weinberg:1971fb}.

As inflation takes place in the large field regime, $\phi \simeq \sqrt{6\alpha}$, it is plausible to assume that the inflationary energy scale is much bigger than the vacuum expectation value of the theory, $\phi \gg v$. Therefore, the quartic power of the non-canonical field in the primordial potential is dominant during inflation.
After the field redefinition (\ref{fieldrelation}), it assumes the form,
\begin{equation}
    V(\varphi) = 9\alpha^2\lambda \left(\tanh{\frac{\varphi}{\sqrt{6\alpha}}}\right)^4,\label{eq:pot_higgs}
\end{equation}
hereafter referred as Higgs-like model. 

Figure \eqref{fig:higgs_sta_pot} shows the primordial potential, as a function of the canonical field $\varphi$, for different values of $\alpha$. A flat regime is obtained for $\varphi \gg \sqrt{6\alpha}$. Such behaviour makes the model's predictions deviate from those of the usual chaotic scenario.  In particular, the ``flattening" of the potential is more efficient for lower values of $\alpha$, while the canonical chaotic inflation $\varphi^4$ is recovered in the limit $\alpha \rightarrow \infty$.

\section{Slow-Roll Analysis}\label{slowrollanalysis}

Inflation and its observables are described in terms of the slow-roll parameters, $\epsilon$ and $\eta$. These parameters can be computed from the inflationary potential by
\begin{eqnarray}
 &\epsilon & = \frac{M^2_{P}}{2}\left(\frac{ V^{\prime}}{ V }\right)^2, \quad \quad
 \eta  = M^2_{P}\frac{V^{\prime \prime}}{V},
 \label{slowparameters}
\end{eqnarray}
where $^\prime$ indicates derivative with respect to $\varphi$. Inflation begins when
$\epsilon,\eta \ll 1$ and comes to an end for $\epsilon,\eta \simeq 1$. In the slow-roll regime, the scalar spectral index and the tensor-to-scalar ratio have the form:
\begin{equation}
    n_s=1-6\epsilon+2\eta \quad \quad{\mbox{and}} \quad \quad r=16\epsilon.
 \label{PParameters}
\end{equation}

We can finally consider the amplitude of the primordial scalar perturbations produced during inflation:
\begin{equation}
    P_R(k_{\ast}) = A_s(k_{\ast}) =\left.\frac{V}{24M^4_P\pi^2\epsilon}\right|_{\varphi=\varphi_\ast},
\label{eq:PR}
\end{equation}
computed for the field strength $\varphi_\ast$ corresponding to the energy scale at which the pivot scale, $k_{\ast}$, crossed the Hubble horizon. The normalization value of $A_s$ is set by the Planck Collaboration to about $2.0933\times 10^{-9}$ for the pivot choice $k_{\ast}=0.05$ Mpc$^{-1}$~\citep{Aghanim:2018eyx}.

The field strength at horizon crossing, $\varphi_\ast$, can be related to the amount of expansion the universe experienced, since the horizon crossing moment up to the end of inflation, via
\begin{equation}
    N = \frac{1}{M^2_P}\int_{\varphi_e}^{\varphi_{\ast}}\frac{V}{V^\prime}d\varphi. \label{efolds}
\end{equation}
Several uncertainties may affect the estimate of the e-fold number. The main source lies from the lack of information about the reheating process, yielding to this quantity a moderate model dependence. For the class of inflationary models which this work is concerned, the number of e-folds can be estimated between $50-60$ \cite{GarciaBellido:2008ab,Akrami:2017cir,Dimopoulos:2017zvq,Turner:1983he,Liddle:2003as,NeferSenoguz:2008nn,Gong:2015qha,Akrami:2020zxw}.

\begin{figure}[ht]
 \begin{center}
 \includegraphics[scale=0.8]{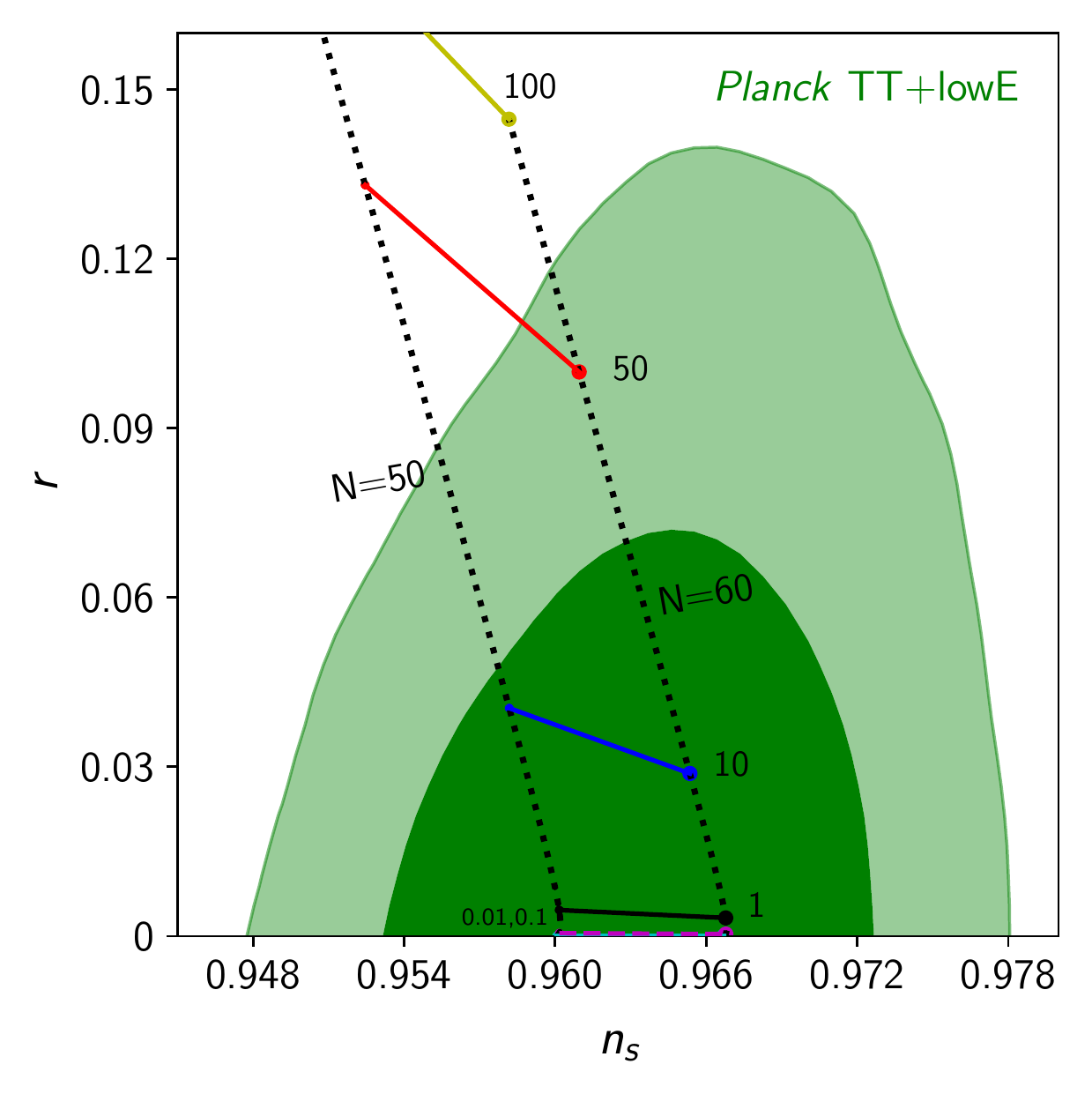}
 \end{center}
\caption{The $n_s - r$ plane for the Higgs-like potential \eqref{eq:pot_higgs}, considering different values of the parameter $\alpha$ \textbf{(solid colored lines and magenta dashed line)} and two values for the number of e-folds, $N=50$ and $N=60$ \textbf{(black dotted lines)}. The contours are the $68\%$ and $95\%$ confidence level regions obtained from Planck (2018) CMB data using the pivot scale $k_{*}=0.05$Mpc$^{-1}$.}
\label{fig:higgs_nsr_TT_theo}
\end{figure}

Once the number of e-folds is set, the expression in \eqref{efolds} can be solved for the field strength at horizon crossing, $\varphi_\ast$. The resulting scale of energy is then employed to compute the model predictions for the inflationary observables. Such procedure is not possible analytically, requiring the use of numerical methods. In particular, we set the number of e-folds in the range $50 \leq N \leq 60$, in order to compare the model predictions with the Planck data.

Figure \eqref{fig:higgs_nsr_TT_theo} shows the behavior of the scalar spectral index and the tensor-to-scalar ratio for the model \eqref{eq:pot_higgs}, considering different values of the parameter $\alpha$ and the number of e-folds ranging from $N=50$ to $N=60$. The results obtained here are in agreement with the ones presented in~\cite{Kallosh:2013yoa}, for the family of $\alpha-$attractors models of type $\phi^n$. In particular, a continuous interpolation between the chaotic inflation at large $\alpha$ and the universal attractor \eqref{AttractorObservable} at small $\alpha$ can be noticed from that figure. For $N=55$, the consistency at $68\%$ confidence level (C.L.) between the model predictions and Planck data occurs for $\alpha \lesssim 28$.

\section{Analysis and Results}\label{analysisandresults}

In order to investigate the observational feasibility of this class of inflationary models, we modify the latest version of the Code for Anisotropies in the Microwave Background (CAMB)~\citep{Lewis_2000} following the lines of the {\sc ModeCode}~\citep{Mortonson:2010er}, adapted to our primordial potential choice (including the parameter $\alpha$). 
On the other hand, to calculate the spectrum of CMB temperature fluctuations from the Primordial Power Spectrum (PPS), we need to solve the equations of the inflationary dynamics, i.e. the Friedmann and Klein-Gordon equations. Additionally, we also need to find the Fourier components associated with curvature perturbations produced by the fluctuations of the scalar field $\varphi$, which can be determined numerically using the {\sc ModeCode}. This code solves the Mukhanov-Sasaki equations~\citep{weinberg2008cosmology}
\begin{eqnarray}
 u''_k+\left(k^2-\frac{z''}{z}\right)u_k=0,
 \end{eqnarray}
 where $u\equiv -z\mathcal{R}$ and $z\equiv a\dot{\varphi}/H$, and $a$, $H$, $\mathcal{R}$ are the scale factor, the Hubble parameter, and the comoving curvature perturbation, respectively. These in turn are used to determine $\mathcal{P_R}(k)$, that is related with $u_k$ and $z$ via:
\begin{eqnarray}
\mathcal{P_R}(k)=\frac{k^3}{2\pi^2}\left|\frac{u_k}{z}\right| ^2.
\end{eqnarray}

\begin{figure}[ht]
 \begin{center}
 \includegraphics[scale=0.3]{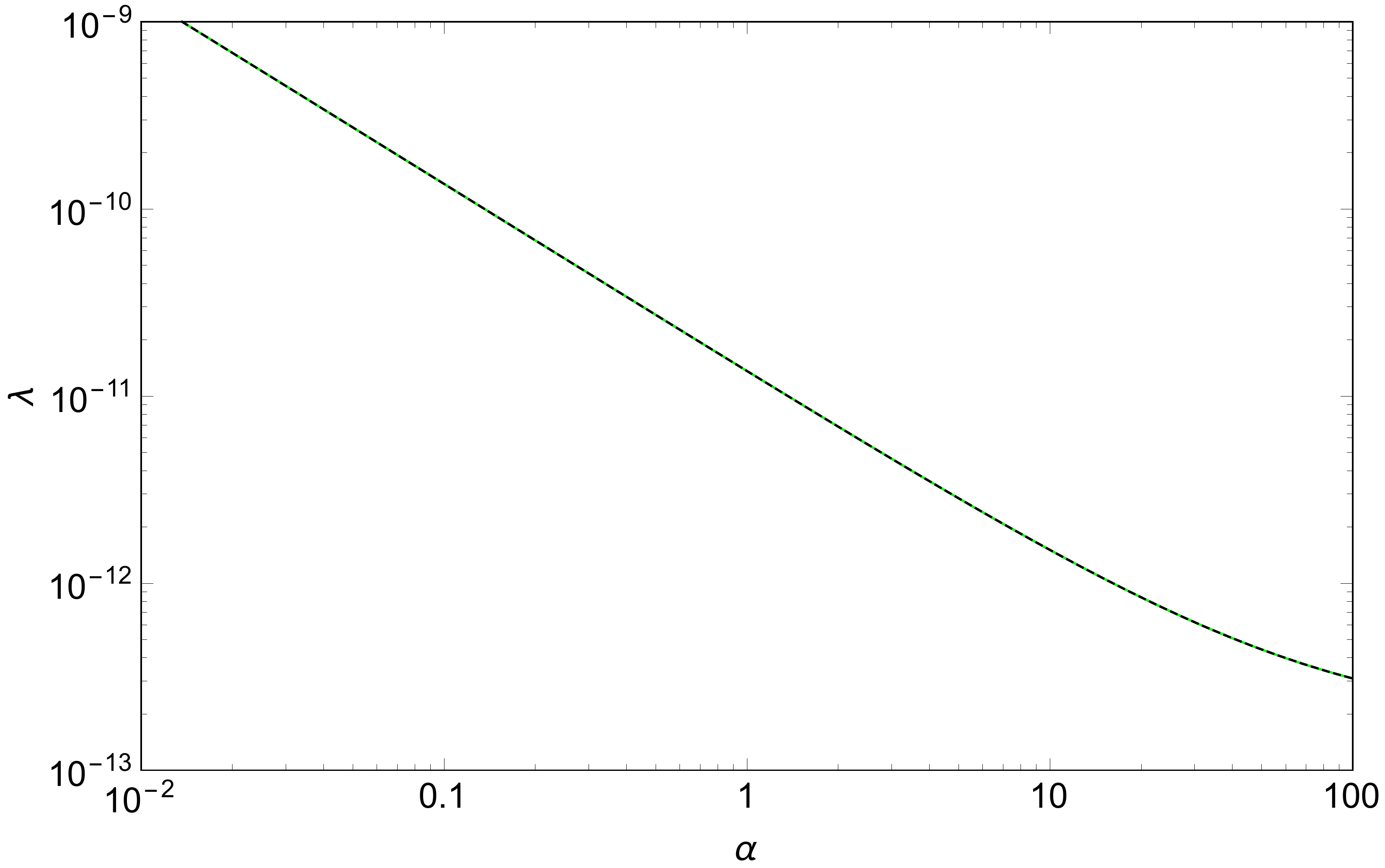}
 \end{center}
\caption{$\lambda$ vs $\alpha$ computed for $N=55$. The green curve interpolates the numerical points while the dashed curve shows the analytical fit to the numerical data points.
}
\label{fig:Lamb_alpha}
\end{figure}

Thus, the method consists in selecting the form of the potential $V(\varphi)$, solve the dynamical equations to obtain $H$ and $\varphi$, and then the PPS is obtained using the solutions of the Mukhanov-Sasaki equations.
As discussed before, we can use Eq.~\eqref{eq:PR}, given the potential $V(\varphi)$ \eqref{eq:pot_higgs}, and invert it in order to determine the potential amplitude $\lambda$ in terms of $\alpha$. In particular, we set $N=55$ to obtain the green curve in figure \eqref{fig:Lamb_alpha}. Note that $\lambda$ decreases with $\alpha$ up to the chaotic $\varphi^4$ value ($\lambda \simeq 10^{-13}$) in the limit $\alpha\rightarrow \infty$. Although an expression of $\lambda(\alpha)$ can not be obtained analytically, one could fit the numerical points to an analytical curve. We adjust the numerical results to the logarithmic power of $\alpha$ up to fifteenth order (represented by the dashed curve in figure \eqref{fig:Lamb_alpha}),

%

\begin{eqnarray}
\lambda(\alpha) \simeq && 1.36\times 10^{-11} - 1.35\times 10^{-11}\ln{\left(\alpha\right)} + 6.80 \times 10^{-12}{\ln{\left(\alpha\right)}}^2 \nonumber \\ &-& 2.28\times 10^{-12}{\ln{\left(\alpha\right)}}^3 + 5.70\times 10^{-13}{\ln{\left(\alpha\right)}}^4 - 1.14\times 10^{-13}{\ln{\left(\alpha\right)}}^5 \nonumber \\ &+& 1.90\times 10^{-14}{\ln{\left(\alpha\right)}}^6 - 2.72 \times 10^{-15}{\ln{\left(\alpha\right)}}^7 + 3.40 \times 10^{-16}{\ln{\left(\alpha\right)}}^8 \nonumber \\ &-& 3.78 \times 10^{-17}{\ln{\left(\alpha\right)}}^9 + 3.80 \times 10^{-18}{\ln{\left(\alpha\right)}}^{10} - 3.40 \times 10^{-19}{\ln{\left(\alpha\right)}}^{11} \nonumber \\ &+& 2.72 \times 10^{-20}{\ln{\left(\alpha\right)}}^{12} - 2.35 \times 10^{-21}{\ln{\left(\alpha\right)}}^{13} + 2.06 \times 10^{-22}{\ln{\left(\alpha\right)}}^{14} \nonumber \\ &-& 9.79 \times 10^{-24}{\ln{\left(\alpha\right)}}^{15}.\label{eq:fit_lambda}
\end{eqnarray}

\begin{figure}[t]
 \begin{center}
 \includegraphics[scale=0.7]{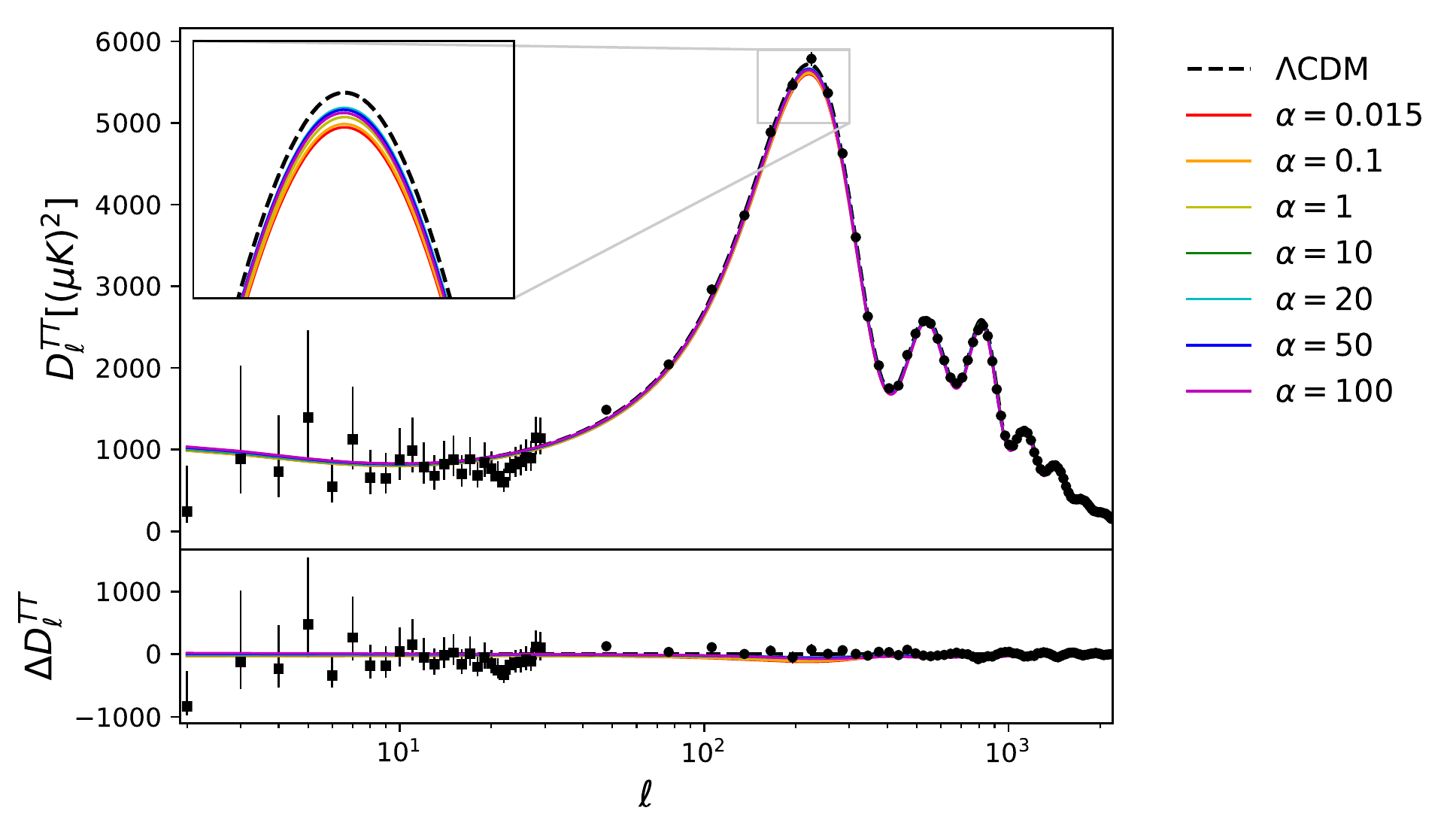}
 \end{center}
\caption{ The theoretical predictions for the CMB temperature angular power spectrum considering different values of $\alpha$ for the Higgs-like model. The data points correspond to the latest release of Planck data~\cite{Aghanim:2018eyx}.}
\label{fig:pspect}
\end{figure}

The theoretical predictions for the Higgs-like potential \eqref{eq:pot_higgs} are shown in Fig.~\eqref{fig:pspect}.
Notice that the effect of $\alpha$ in the temperature angular power spectrum is a slight variation in its amplitude. However, this variation does not follow a correlation with the increment of $\alpha$, instead, the amplitude oscillates to upper and lower values arbitrarily.
In addition, despite the $n_s - r$ plan has shown that the value of $\alpha=100$ is out of the $95\%$ confidence region allowed by the data (for both $N=50$ and $N=60$), the theoretical predictions are close to the standard $\Lambda$CDM model. Therefore, we consider as an appropriate range for our analysis the flat prior of $0.01 < \alpha < 100$.

To explore the cosmological parameter space of the Higgs-like model and investigate the impact of the parameter $\alpha$, we perform a Markov Chain Monte Carlo (MCMC) analysis, using the most recent version of {\sc CosmoMC} code~\citep{Lewis_2002}. We also vary the usual cosmological parameters, namely, the baryon and the cold dark matter density, the ratio between the sound horizon and the angular diameter distance at decoupling, and the optical depth: $\left \{\Omega_bh^2~,~\Omega_ch^2~,~\theta~,~\tau\right \}$. We consider purely adiabatic initial conditions, fix the sum of neutrino masses to $0.06~eV$ and the universe curvature to zero, and also vary the nuisance foregrounds parameters~\citep{Aghanim:2015xee}.
The priors we considered on the cosmological parameters are shown in Table \eqref{tab_priors}. 

 \begin{table}
 \centering
 \caption{Priors on the cosmological parameters considered in the analysis.}
 {\begin{tabular}{|c|c|}
 \hline 
 Parameter & Prior Ranges \\ 
 \hline
$\Omega_{b}h^{2}$ & $[0.005 : 0.1]$ \\ 
$\Omega_{c}h^{2}$ & $[0.001 : 0.99]$ \\ 
$\theta$ & $[0.5 : 10.0]$ \\ 
$\tau$ & $[0.01 : 0.8]$ \\ 
 
 
$\alpha$ & $[0.01 : 100]$ \\ 
\hline 
\end{tabular}\label{tab_priors}}
\end{table} 

\begin{table}[t]
\centering
\caption{$68\%$ confidence constraints for the cosmological parameters of the Higgs-like model using PLA18+BAO+BK15 data.
The table is divided into two sections: the upper section shows the primary parameters, while in the lower part shows the derived ones.}
\scalebox{1}{
{\begin{tabular}{|c|c|c|}
\hline
&\multicolumn{2}{c|}{Higgs-like} \\
\hline
 {Parameter} & {mean} & {best fit} \\
\hline
Primary & & \\
$\Omega_b h^2$
& $0.02212 \pm 0.018$ 
& $0.02208$
\\
$\Omega_{c} h^2$
& $0.1199 \pm 0.0009$ 
& $0.1196$
\\
$\theta$
& $1.04085 \pm 0.00040$ 
& $1.04108$
\\
$\tau$
& $0.048 \pm 0.003$ 
& $0.048$
\\
$\alpha$
& $7.56 \pm 5.15$
& $9.82$
\\
\hline
\hline
Derived & & \\
$H_0$
& $67.16 \pm 0.38$ 
& $67.21$
\\
{$\Omega_{m}$}
& $0.316 \pm  0.005$ 
& $0.320$
\\
{$\Omega_{\Lambda}$}
& $0.684 \pm 0.005$ 
& $0.680$
\\
{$n_s$}
& $ 0.963 \pm 0.001$ 
& $0.9621$
\\
{$r_{0.002}$}
& $0.023\pm 0.014$
& $0.030$
\\
\hline
\end{tabular} \label{tab:Tabel_results_1}}
}
\end{table} 

In order to compare the theoretical predictions of the model with the observational data, we use the CMB data from the latest release of Planck Collaboration (2018)~\citep{collaboration2019planck}. We consider high multipoles Planck temperature data from the 100-,143-, and 217-GHz half-mission T maps, and  the low multipoles data by the joint TT, EE, BB and TE likelihood, where EE and BB are the E- and B-mode CMB polarization power spectrum and TE is the cross-correlation temperature-polarization (hereafter ``PLA18"). 
In addition, we consider an extended dataset combining the CMB data with Baryon Acoustic Oscillations (BAO) coming from the 6dF Galaxy Survey (6dFGS)~\citep{bao1}, Sloan Digital Sky Survey (SDSS) DR7 Main Galaxy Sample galaxies~\citep{bao2}, BOSSgalaxy samples, LOWZ and CMASS~\citep{bao3}, and the tensor amplitude of B-mode polarization from the Keck Array and BICEP2 Collaborations~\citep{bicep21,bicep22}, using the BICEP2/Keck field, from 95, 150, and 220 GHz maps (hereafter ``BK15'').

\begin{figure}[]
\centering
	\includegraphics[scale=0.4]{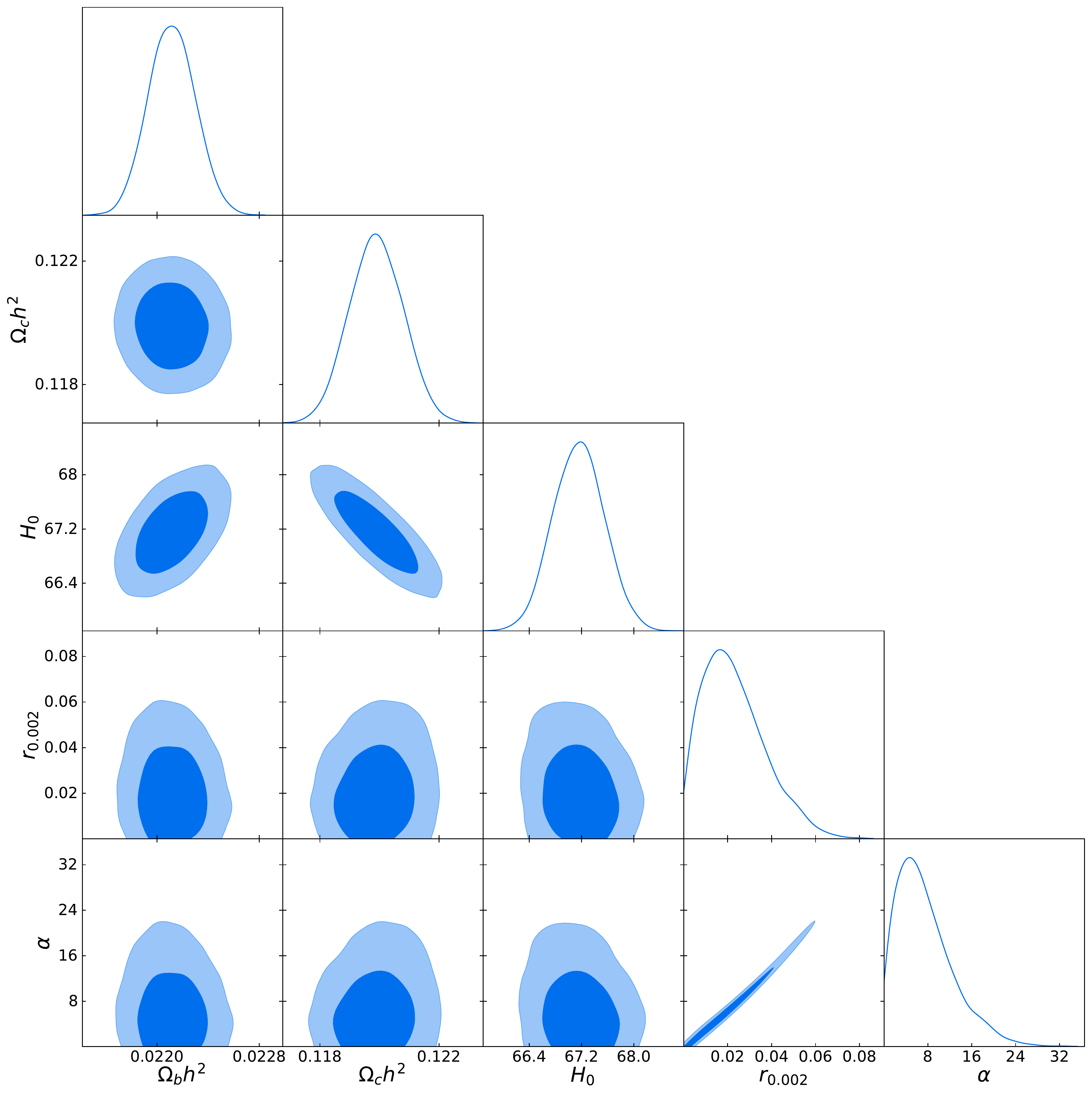}
	\caption{$68\%$ and $95\%$ confidence regions for the primary and derived parameters of the Higgs-like model.}
	\label{fig:higgs_triplot}
\end{figure}

The observational constraints obtained for the Higgs-like potential, using the Planck 2018 likelihood jointly with BAO and BICEP/Keck 2015 data, are summarized in Table \eqref{tab:Tabel_results_1} and in Figure \eqref{fig:higgs_triplot}. Table \eqref{tab:Tabel_results_1} shows the constraints on the cosmological parameters for Higgs-like model. Note that both primary and derived cosmological parameters of Higgs-like model are in good agreement with $\Lambda$CDM within $1\sigma$ level~\cite{Aghanim:2018eyx}. From this analysis, we find $\alpha=7.56\pm 5.15$ at $68\%$ C.L. and at $95\%$ C.L. we find an upper limit of $\alpha<17.69$, that is considerably more stringent than the slow-roll result $\alpha \lesssim 28$. Other recent analysis is found in \cite{Akrami:2017cir,Garcia-Garcia:2018hlc,Cedeno:2019cgr}.

\begin{figure}[ht]
 \begin{center}
 \includegraphics[scale=0.7]{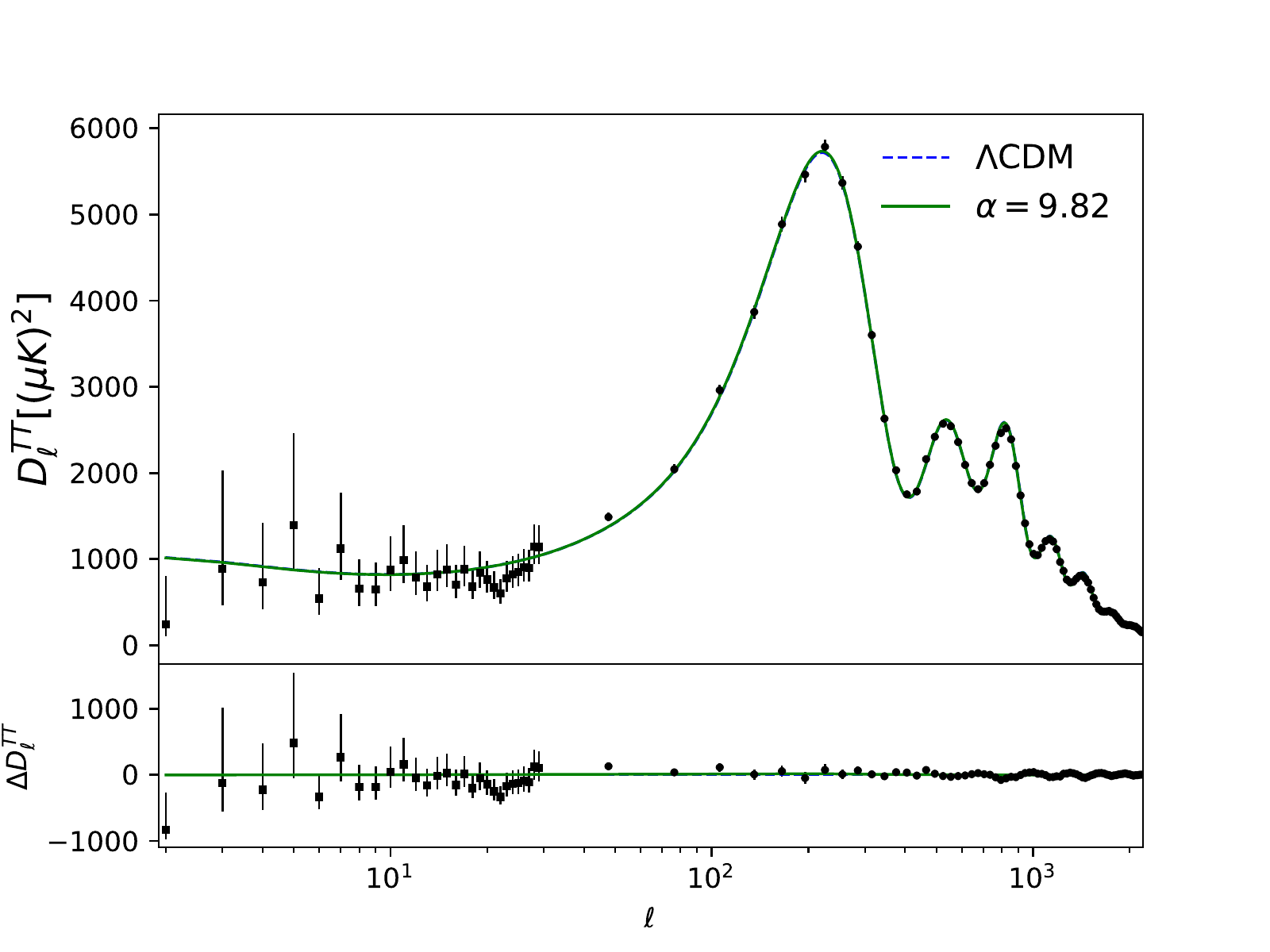}
  \end{center}
\caption{The best-fit CMB temperature angular power spectra for the Higgs-like model (green curve) and $\Lambda$CDM model (blue curve). The data points correspond to the latest release of Planck (2018) data and the lower panel show the residuals with respect to the reference model ($\Lambda$CDM).}
\label{fig:higgs_triplot2}
\end{figure}

Figure \eqref{fig:higgs_triplot} shows the confidence regions at $68\%$ and $95\%$ C.L. and the posterior probability distribution for the main cosmological parameters of the Higgs-like model. The bounds obtained for the matter density and Hubble parameter are in accordance with $\Lambda$CDM model~\citep{Aghanim:2018eyx}. However, note that the limits on the tensor-to-scalar ratio, an upper bound of $r<0.049$ at 95\% C.L., are tighter than that obtained previously for both $\Lambda$CDM model and $\alpha-$attractor models~\citep{Aghanim:2018eyx,Kallosh:2013yoa,Kallosh_2015,Kallosh_2014}. Also, there is a positive correlation between $\alpha$ and $r$. Indeed, this fact can also be inferred from the $n_s - r$ plan shown in Fig.~\eqref{fig:higgs_nsr_TT_theo}, which exhibits this behavior between $\alpha$ and $r$. Lastly, the figure \eqref{fig:higgs_triplot2} displays the CMB temperature angular power spectrum for the best-fit values of the Higgs-like model, and shows a fit to the CMB data as good as the one provided by $\Lambda$CDM model.

%

\section{Discussion and Conclusions}\label{conclusions}

In this paper, we analyzed a class of inflationary models based on supergravity theories, namely, $\alpha-$attractors. Specifically, we investigated the observational viability of the model for the double-well inflationary potential. Such choice is strongly motivated in the context of particle physics, presenting an ideal framework not only to successfully carry out inflation but also to connect the inflationary dynamics to low-energy phenomenology of fundamental particles. For this reason, the resulting model was called Higgs-like model.

For large values of $\alpha$ ($\mathcal{O}(10^2)$), the slow-roll analysis revealed that the model predictions are outside the region allowed by the data on the $n_s -  r$ plan, as seen in Fig.~\eqref{fig:higgs_nsr_TT_theo}. As $\alpha$ decreases, the predictions converge to the universal attractor values, Eq.~\eqref{AttractorObservable}. Such behavior is similar to the one predicted by a general family of cosmological attractors with $V(\varphi)\sim \varphi^n$~\citep{Kallosh:2013yoa}. 
Extending the slow-roll study, we performed a MCMC analysis in order to explore the parameter space. The results found for the primary and derived cosmological parameters show an excellent match to the latest cosmological data as well as with the predictions of the $\Lambda$CDM model.


We also obtained a tight constraint on the parameter $\alpha$ using an extended dataset including CMB and BAO data, i.e., $\alpha=7.56\pm 5.15$ ($68\%$ C.L.).
Particularly from maximal supergravity, string theory, and M-theory, there are seven preferred values for the $\alpha$ parameter, $3\alpha =1,2,...,7$ \cite{Akrami:2017cir,Ferrara:2016fwe,Kallosh:2017wnt,Kallosh:2017ced}, with all of them being favoured by current data (according to $\alpha<17.69$ at $95\%$ C.L. upper limit). Also, the best-fit value, $\alpha = 9.82$, moves the prediction of the inflationary observables, $n_s$ and $r$, farther away from the universal attractor value given by Eq.~\eqref{AttractorObservable}. This may be suggesting a substantial difference in the predictive pattern of the $\alpha$-attractor models for different inflationary potentials, $V(\phi)$, {{in contrast to what is expected from the usual slow-roll analysis \cite{Kallosh:2013yoa,Linde:2016uec}}}.

It is worth mentioning that a similar analysis was also performed considering the Starobinsky potential (\ref{StarLike})  in the context of $\alpha-$attractor models. The results showed no significant improvement of the constraints on the cosmological parameters with the bounds on $\alpha$ being not restrictive, which confirms the robustness of the Starobinsky model, as recently discussed in \cite{Renzi:2019ewp,SantosdaCosta:2020ont}. 

In conclusion, we attested the observational viability of the Higgs-like attractor inflation in light of current observational data. However, further investigation about the dependence of those predictions on the choice of non-canonical potential $V(\phi)$ is needed. We shall consider this question in a forthcoming communication.

\section*{Acknowledgements}

J.~G. Rodrigues acknowledges Conselho Nacional de Desenvolvimento Cient\'{\i}fico e Tecnol\'ogico (CNPq) for the financial support. 
S.~Santos da Costa thanks the financial support from the Programa de Capacita\c{c}\~ao Institucional (PCI) do Observat\'orio Nacional/MCTI. J.~Alcaniz is supported CNPq (Grants no. 310790/2014-0 and 400471/2014-0) and Funda\c{c}\~ao de Amparo \`a Pesquisa do Estado do Rio de Janeiro FAPERJ (grant no. 233906). We also thank the authors of the ModeCode (M.~Mortonson, H.
~Peiris and R.~Easther) and CosmoMC (A.~Lewis) codes, and the computational support of the Observat\'orio Nacional Data Center.

\vspace{2cm}



\bibliographystyle{elsarticle-num} 
\bibliography{bibliografia}

\begin{thebibliography}{10}
\expandafter\ifx\csname url\endcsname\relax
  \def\url#1{\texttt{#1}}\fi
\expandafter\ifx\csname urlprefix\endcsname\relax\def\urlprefix{URL }\fi
\expandafter\ifx\csname href\endcsname\relax
  \def\href#1#2{#2} \def\path#1{#1}\fi

\bibitem{Guth:1980zm}
A.~H. Guth, {The Inflationary Universe: A Possible Solution to the Horizon and
  Flatness Problems}, Phys. Rev. D23 (1981) 347--356.
\newblock \href {https://doi.org/10.1103/PhysRevD.23.347}
  {\path{doi:10.1103/PhysRevD.23.347}}.

\bibitem{Linde:1981mu}
A.~D. Linde, {A New Inflationary Universe Scenario: A Possible Solution of the
  Horizon, Flatness, Homogeneity, Isotropy and Primordial Monopole Problems},
  Phys. Lett. 108B (1982) 389--393.
\newblock \href {https://doi.org/10.1016/0370-2693(82)91219-9}
  {\path{doi:10.1016/0370-2693(82)91219-9}}.

\bibitem{Albrecht:1982wi}
A.~Albrecht, P.~J. Steinhardt, {Cosmology for Grand Unified Theories with
  Radiatively Induced Symmetry Breaking}, Phys. Rev. Lett. 48 (1982)
  1220--1223.
\newblock \href {https://doi.org/10.1103/PhysRevLett.48.1220}
  {\path{doi:10.1103/PhysRevLett.48.1220}}.

\bibitem{Martin:2013tda}
J.~Martin, C.~Ringeval, V.~Vennin, {Encyclop\ae dia Inflationaris}, Phys. Dark
  Univ. 5-6 (2014) 75--235.
\newblock \href {http://arxiv.org/abs/1303.3787} {\path{arXiv:1303.3787}},
  \href {https://doi.org/10.1016/j.dark.2014.01.003}
  {\path{doi:10.1016/j.dark.2014.01.003}}.

\bibitem{Starobinsky:1980te}
A.~A. Starobinsky, {A New Type of Isotropic Cosmological Models Without
  Singularity}, Phys. Lett. 91B (1980) 99--102, [,771(1980)].
\newblock \href {https://doi.org/10.1016/0370-2693(80)90670-X}
  {\path{doi:10.1016/0370-2693(80)90670-X}}.

\bibitem{mukhanov1981quantum}
V.~F. Mukhanov, G.~Chibisov, Quantum fluctuations and a nonsingular universe,
  JETP Letters 33~(10) (1981) 532--535.

\bibitem{Whitt:1984pd}
B.~Whitt, {Fourth Order Gravity as General Relativity Plus Matter}, Phys. Lett.
  B 145 (1984) 176--178.
\newblock \href {https://doi.org/10.1016/0370-2693(84)90332-0}
  {\path{doi:10.1016/0370-2693(84)90332-0}}.

\bibitem{Salopek:1988qh}
D.~S. Salopek, J.~R. Bond, J.~M. Bardeen, {Designing Density Fluctuation
  Spectra in Inflation}, Phys. Rev. D40 (1989) 1753.
\newblock \href {https://doi.org/10.1103/PhysRevD.40.1753}
  {\path{doi:10.1103/PhysRevD.40.1753}}.

\bibitem{Fakir:1990eg}
R.~Fakir, W.~G. Unruh, {Improvement on cosmological chaotic inflation through
  nonminimal coupling}, Phys. Rev. D41 (1990) 1783--1791.
\newblock \href {https://doi.org/10.1103/PhysRevD.41.1783}
  {\path{doi:10.1103/PhysRevD.41.1783}}.

\bibitem{Makino:1991sg}
N.~Makino, M.~Sasaki, {The Density perturbation in the chaotic inflation with
  nonminimal coupling}, Prog. Theor. Phys. 86 (1991) 103--118.
\newblock \href {https://doi.org/10.1143/PTP.86.103}
  {\path{doi:10.1143/PTP.86.103}}.

\bibitem{Kallosh:2013hoa}
R.~Kallosh, A.~Linde, {Universality Class in Conformal Inflation}, JCAP 1307
  (2013) 002.
\newblock \href {http://arxiv.org/abs/1306.5220} {\path{arXiv:1306.5220}},
  \href {https://doi.org/10.1088/1475-7516/2013/07/002}
  {\path{doi:10.1088/1475-7516/2013/07/002}}.

\bibitem{Kallosh:2013tua}
R.~Kallosh, A.~Linde, D.~Roest, {Universal Attractor for Inflation at Strong
  Coupling}, Phys. Rev. Lett. 112~(1) (2014) 011303.
\newblock \href {http://arxiv.org/abs/1310.3950} {\path{arXiv:1310.3950}},
  \href {https://doi.org/10.1103/PhysRevLett.112.011303}
  {\path{doi:10.1103/PhysRevLett.112.011303}}.

\bibitem{Kallosh:2013yoa}
R.~Kallosh, A.~Linde, D.~Roest, {Superconformal Inflationary
  $\alpha$-Attractors}, JHEP 11 (2013) 198.
\newblock \href {http://arxiv.org/abs/1311.0472} {\path{arXiv:1311.0472}},
  \href {https://doi.org/10.1007/JHEP11(2013)198}
  {\path{doi:10.1007/JHEP11(2013)198}}.

\bibitem{Akrami:2018odb}
Y.~Akrami, et~al., {Planck 2018 results. X. Constraints on inflation}, Astron.
  Astrophys. 641 (2020) A10.
\newblock \href {http://arxiv.org/abs/1807.06211} {\path{arXiv:1807.06211}},
  \href {https://doi.org/10.1051/0004-6361/201833887}
  {\path{doi:10.1051/0004-6361/201833887}}.

\bibitem{Martin:2013nzq}
J.~Martin, C.~Ringeval, R.~Trotta, V.~Vennin, {The Best Inflationary Models
  After Planck}, JCAP 03 (2014) 039.
\newblock \href {http://arxiv.org/abs/1312.3529} {\path{arXiv:1312.3529}},
  \href {https://doi.org/10.1088/1475-7516/2014/03/039}
  {\path{doi:10.1088/1475-7516/2014/03/039}}.

\bibitem{Linde_2011}
A.~Linde, M.~Noorbala, A.~Westphal, Observational consequences of chaotic
  inflation with nonminimal coupling to gravity, Journal of Cosmology and
  Astroparticle Physics 2011~(03) (2011) 013–013.

\bibitem{Kehagias:2013mya}
A.~Kehagias, A.~Moradinezhad~Dizgah, A.~Riotto, {Remarks on the Starobinsky
  model of inflation and its descendants}, Phys. Rev. D 89~(4) (2014) 043527.
\newblock \href {http://arxiv.org/abs/1312.1155} {\path{arXiv:1312.1155}},
  \href {https://doi.org/10.1103/PhysRevD.89.043527}
  {\path{doi:10.1103/PhysRevD.89.043527}}.

\bibitem{Aghanim:2018eyx}
N.~Aghanim, et~al., {Planck 2018 results. VI. Cosmological parameters} (2018).
\newblock \href {http://arxiv.org/abs/[1807.06209]}
  {\path{arXiv:[1807.06209]}}.

\bibitem{Bezrukov:2007ep}
F.~L. Bezrukov, M.~Shaposhnikov, {The Standard Model Higgs boson as the
  inflaton}, Phys. Lett. B659 (2008) 703--706.
\newblock \href {http://arxiv.org/abs/0710.3755} {\path{arXiv:0710.3755}},
  \href {https://doi.org/10.1016/j.physletb.2007.11.072}
  {\path{doi:10.1016/j.physletb.2007.11.072}}.

\bibitem{Barvinsky:2008ia}
A.~O. Barvinsky, A.~{\relax Yu}. Kamenshchik, A.~A. Starobinsky, {Inflation
  scenario via the Standard Model Higgs boson and LHC}, JCAP 0811 (2008) 021.
\newblock \href {http://arxiv.org/abs/0809.2104} {\path{arXiv:0809.2104}},
  \href {https://doi.org/10.1088/1475-7516/2008/11/021}
  {\path{doi:10.1088/1475-7516/2008/11/021}}.

\bibitem{GarciaBellido:2008ab}
J.~Garcia-Bellido, D.~G. Figueroa, J.~Rubio, {Preheating in the Standard Model
  with the Higgs-Inflaton coupled to gravity}, Phys. Rev. D79 (2009) 063531.
\newblock \href {http://arxiv.org/abs/0812.4624} {\path{arXiv:0812.4624}},
  \href {https://doi.org/10.1103/PhysRevD.79.063531}
  {\path{doi:10.1103/PhysRevD.79.063531}}.

\bibitem{Bezrukov:2014bra}
F.~Bezrukov, M.~Shaposhnikov, {Higgs inflation at the critical point}, Phys.
  Lett. B734 (2014) 249--254.
\newblock \href {http://arxiv.org/abs/1403.6078} {\path{arXiv:1403.6078}},
  \href {https://doi.org/10.1016/j.physletb.2014.05.074}
  {\path{doi:10.1016/j.physletb.2014.05.074}}.

\bibitem{Hamada:2014iga}
Y.~Hamada, H.~Kawai, K.-y. Oda, S.~C. Park, {Higgs Inflation is Still Alive
  after the Results from BICEP2}, Phys. Rev. Lett. 112~(24) (2014) 241301.
\newblock \href {http://arxiv.org/abs/1403.5043} {\path{arXiv:1403.5043}},
  \href {https://doi.org/10.1103/PhysRevLett.112.241301}
  {\path{doi:10.1103/PhysRevLett.112.241301}}.

\bibitem{Gomes:2016cwj}
C.~Gomes, J.~G. Rosa, O.~Bertolami, {Inflation in non-minimal matter-curvature
  coupling theories}, JCAP 06 (2017) 021.
\newblock \href {http://arxiv.org/abs/1611.02124} {\path{arXiv:1611.02124}},
  \href {https://doi.org/10.1088/1475-7516/2017/06/021}
  {\path{doi:10.1088/1475-7516/2017/06/021}}.

\bibitem{Lee:2018esk}
H.~M. Lee, {Light inflaton completing Higgs inflation}, Phys. Rev. D98~(1)
  (2018) 015020.
\newblock \href {http://arxiv.org/abs/1802.06174} {\path{arXiv:1802.06174}},
  \href {https://doi.org/10.1103/PhysRevD.98.015020}
  {\path{doi:10.1103/PhysRevD.98.015020}}.

\bibitem{Lerner:2009xg}
R.~N. Lerner, J.~McDonald, {Gauge singlet scalar as inflaton and thermal relic
  dark matter}, Phys. Rev. D80 (2009) 123507.
\newblock \href {http://arxiv.org/abs/0909.0520} {\path{arXiv:0909.0520}},
  \href {https://doi.org/10.1103/PhysRevD.80.123507}
  {\path{doi:10.1103/PhysRevD.80.123507}}.

\bibitem{Okada:2011en}
N.~Okada, M.~U. Rehman, Q.~Shafi, {Non-Minimal B-L Inflation with Observable
  Gravity Waves}, Phys. Lett. B701 (2011) 520--525.
\newblock \href {http://arxiv.org/abs/1102.4747} {\path{arXiv:1102.4747}},
  \href {https://doi.org/10.1016/j.physletb.2011.06.044}
  {\path{doi:10.1016/j.physletb.2011.06.044}}.

\bibitem{Ballesteros:2016euj}
G.~Ballesteros, J.~Redondo, A.~Ringwald, C.~Tamarit, {Unifying inflation with
  the axion, dark matter, baryogenesis and the seesaw mechanism}, Phys. Rev.
  Lett. 118~(7) (2017) 071802.
\newblock \href {http://arxiv.org/abs/1608.05414} {\path{arXiv:1608.05414}},
  \href {https://doi.org/10.1103/PhysRevLett.118.071802}
  {\path{doi:10.1103/PhysRevLett.118.071802}}.

\bibitem{Ferreira:2017ynu}
J.~G. Ferreira, C.~A. de~S.~Pires, J.~G. Rodrigues, P.~S. Rodrigues~da Silva,
  {Inflation scenario driven by a low energy physics inflaton}, Phys. Rev.
  D96~(10) (2017) 103504.
\newblock \href {http://arxiv.org/abs/1707.01049} {\path{arXiv:1707.01049}},
  \href {https://doi.org/10.1103/PhysRevD.96.103504}
  {\path{doi:10.1103/PhysRevD.96.103504}}.

\bibitem{Kallosh:2014rga}
R.~Kallosh, A.~Linde, D.~Roest, {Large field inflation and double
  $\alpha$-attractors}, JHEP 08 (2014) 052.
\newblock \href {http://arxiv.org/abs/1405.3646} {\path{arXiv:1405.3646}},
  \href {https://doi.org/10.1007/JHEP08(2014)052}
  {\path{doi:10.1007/JHEP08(2014)052}}.

\bibitem{Kallosh:2015lwa}
R.~Kallosh, A.~Linde, {Planck, LHC, and $\alpha$-attractors}, Phys. Rev. D 91
  (2015) 083528.
\newblock \href {http://arxiv.org/abs/1502.07733} {\path{arXiv:1502.07733}},
  \href {https://doi.org/10.1103/PhysRevD.91.083528}
  {\path{doi:10.1103/PhysRevD.91.083528}}.

\bibitem{Roest:2015qya}
D.~Roest, M.~Scalisi, {Cosmological attractors from $\alpha$-scale
  supergravity}, Phys. Rev. D 92 (2015) 043525.
\newblock \href {http://arxiv.org/abs/1503.07909} {\path{arXiv:1503.07909}},
  \href {https://doi.org/10.1103/PhysRevD.92.043525}
  {\path{doi:10.1103/PhysRevD.92.043525}}.

\bibitem{Linde:2016uec}
A.~Linde, {Random Potentials and Cosmological Attractors}, JCAP 1702~(02)
  (2017) 028.
\newblock \href {http://arxiv.org/abs/1612.04505} {\path{arXiv:1612.04505}},
  \href {https://doi.org/10.1088/1475-7516/2017/02/028}
  {\path{doi:10.1088/1475-7516/2017/02/028}}.

\bibitem{Terada:2016nqg}
T.~Terada, {Generalized Pole Inflation: Hilltop, Natural, and Chaotic
  Inflationary Attractors}, Phys. Lett. B 760 (2016) 674--680.
\newblock \href {http://arxiv.org/abs/1602.07867} {\path{arXiv:1602.07867}},
  \href {https://doi.org/10.1016/j.physletb.2016.07.058}
  {\path{doi:10.1016/j.physletb.2016.07.058}}.

\bibitem{Ueno:2016dim}
Y.~Ueno, K.~Yamamoto, {Constraints on $\alpha$-attractor inflation and
  reheating}, Phys. Rev. D 93~(8) (2016) 083524.
\newblock \href {http://arxiv.org/abs/1602.07427} {\path{arXiv:1602.07427}},
  \href {https://doi.org/10.1103/PhysRevD.93.083524}
  {\path{doi:10.1103/PhysRevD.93.083524}}.

\bibitem{Odintsov:2016vzz}
S.~Odintsov, V.~Oikonomou, {Inflationary $\alpha$-attractors from $F(R)$
  gravity}, Phys. Rev. D 94~(12) (2016) 124026.
\newblock \href {http://arxiv.org/abs/1612.01126} {\path{arXiv:1612.01126}},
  \href {https://doi.org/10.1103/PhysRevD.94.124026}
  {\path{doi:10.1103/PhysRevD.94.124026}}.

\bibitem{Akrami:2017cir}
Y.~Akrami, R.~Kallosh, A.~Linde, V.~Vardanyan, {Dark energy,
  $\alpha$-attractors, and large-scale structure surveys}, JCAP 06 (2018) 041.
\newblock \href {http://arxiv.org/abs/1712.09693} {\path{arXiv:1712.09693}},
  \href {https://doi.org/10.1088/1475-7516/2018/06/041}
  {\path{doi:10.1088/1475-7516/2018/06/041}}.

\bibitem{Dimopoulos:2017zvq}
K.~Dimopoulos, C.~Owen, {Quintessential Inflation with $\alpha$-attractors},
  JCAP 06 (2017) 027.
\newblock \href {http://arxiv.org/abs/1703.00305} {\path{arXiv:1703.00305}},
  \href {https://doi.org/10.1088/1475-7516/2017/06/027}
  {\path{doi:10.1088/1475-7516/2017/06/027}}.

\bibitem{Pozdeeva:2020shl}
E.~O. Pozdeeva, {Generalization of cosmological attractor approach to
  Einstein--Gauss--Bonnet gravity}, Eur. Phys. J. C 80~(7) (2020) 612.
\newblock \href {http://arxiv.org/abs/2005.10133} {\path{arXiv:2005.10133}},
  \href {https://doi.org/10.1140/epjc/s10052-020-8176-3}
  {\path{doi:10.1140/epjc/s10052-020-8176-3}}.

\bibitem{Odintsov:2020thl}
S.~Odintsov, V.~Oikonomou, {Inflationary attractors in $F(R)$ gravity}, Phys.
  Lett. B 807 (2020) 135576.
\newblock \href {http://arxiv.org/abs/2005.12804} {\path{arXiv:2005.12804}},
  \href {https://doi.org/10.1016/j.physletb.2020.135576}
  {\path{doi:10.1016/j.physletb.2020.135576}}.

\bibitem{Garcia-Garcia:2018hlc}
C.~Garc\'\i{}a-Garc\'\i{}a, E.~V. Linder, P.~Ru\'\i{}z-Lapuente,
  M.~Zumalac\'arregui, {Dark energy from $\alpha$-attractors: phenomenology and
  observational constraints}, JCAP 08 (2018) 022.
\newblock \href {http://arxiv.org/abs/1803.00661} {\path{arXiv:1803.00661}},
  \href {https://doi.org/10.1088/1475-7516/2018/08/022}
  {\path{doi:10.1088/1475-7516/2018/08/022}}.

\bibitem{Cedeno:2019cgr}
F.~X. Linares Cede\~no, A.~Montiel, J.~C. Hidalgo, G.~Germ\'an, {Bayesian
  evidence for $\alpha$-attractor dark energy models}, JCAP 08 (2019) 002.
\newblock \href {http://arxiv.org/abs/1905.00834} {\path{arXiv:1905.00834}},
  \href {https://doi.org/10.1088/1475-7516/2019/08/002}
  {\path{doi:10.1088/1475-7516/2019/08/002}}.

\bibitem{collaboration2019planck}
P.~Collaboration, N.~A. et~al., Planck 2018 results. v. cmb power spectra and
  likelihoods (2019).
\newblock \href {http://arxiv.org/abs/1907.12875} {\path{arXiv:1907.12875}}.

\bibitem{bao1}
F.~Beutler, C.~Blake, M.~Colless, D.~H. Jones, L.~Staveley-Smith, L.~Campbell,
  Q.~Parker, W.~Saunders, F.~Watson, {The 6dF Galaxy Survey: baryon acoustic
  oscillations and the local Hubble constant}, Monthly Notices of the Royal
  Astronomical Society 416~(4) (2011) 3017--3032.

\bibitem{bao2}
A.~J. Ross, L.~Samushia, C.~Howlett, W.~J. Percival, A.~Burden, M.~Manera, {The
  clustering of the SDSS DR7 main Galaxy sample – I. A 4 per cent distance
  measure at z = 0.15}, Monthly Notices of the Royal Astronomical Society
  449~(1) (2015) 835--847.

\bibitem{bao3}
L.~e.~a. Anderson, {The clustering of galaxies in the SDSS-III Baryon
  Oscillation Spectroscopic Survey: baryon acoustic oscillations in the Data
  Releases 10 and 11 Galaxy samples}, Monthly Notices of the Royal Astronomical
  Society 441~(1) (2014) 24--62.

\bibitem{bicep21}
P.~A. R. e.~a. Ade, Joint analysis of bicep2/keck array and planck data, Phys.
  Rev. Lett. 114 (2015) 101301.

\bibitem{bicep22}
P.~A. R. e.~a. Ade, Improved constraints on cosmology and foregrounds from
  bicep2 and keck array cosmic microwave background data with inclusion of 95
  ghz band, Phys. Rev. Lett. 116 (2016) 031302.

\bibitem{Goldstone:1961eq}
J.~Goldstone, {Field Theories with Superconductor Solutions}, Nuovo Cim. 19
  (1961) 154--164.
\newblock \href {https://doi.org/10.1007/BF02812722}
  {\path{doi:10.1007/BF02812722}}.

\bibitem{Higgs:1964ia}
P.~W. Higgs, {Broken symmetries, massless particles and gauge fields}, Phys.
  Lett. 12 (1964) 132--133.
\newblock \href {https://doi.org/10.1016/0031-9163(64)91136-9}
  {\path{doi:10.1016/0031-9163(64)91136-9}}.

\bibitem{Higgs:1966ev}
P.~W. Higgs, {Spontaneous Symmetry Breakdown without Massless Bosons}, Phys.
  Rev. 145 (1966) 1156--1163.
\newblock \href {https://doi.org/10.1103/PhysRev.145.1156}
  {\path{doi:10.1103/PhysRev.145.1156}}.

\bibitem{Weinberg:1967tq}
S.~Weinberg, {A Model of Leptons}, Phys. Rev. Lett. 19 (1967) 1264--1266.
\newblock \href {https://doi.org/10.1103/PhysRevLett.19.1264}
  {\path{doi:10.1103/PhysRevLett.19.1264}}.

\bibitem{Weinberg:1971fb}
S.~Weinberg, {Physical Processes in a Convergent Theory of the Weak and
  Electromagnetic Interactions}, Phys. Rev. Lett. 27 (1971) 1688--1691.
\newblock \href {https://doi.org/10.1103/PhysRevLett.27.1688}
  {\path{doi:10.1103/PhysRevLett.27.1688}}.

\bibitem{Turner:1983he}
M.~S. Turner, {Coherent Scalar Field Oscillations in an Expanding Universe},
  Phys. Rev. D 28 (1983) 1243.
\newblock \href {https://doi.org/10.1103/PhysRevD.28.1243}
  {\path{doi:10.1103/PhysRevD.28.1243}}.

\bibitem{Liddle:2003as}
A.~R. Liddle, S.~M. Leach, {How long before the end of inflation were
  observable perturbations produced?}, Phys. Rev. D68 (2003) 103503.
\newblock \href {http://arxiv.org/abs/astro-ph/0305263}
  {\path{arXiv:astro-ph/0305263}}, \href
  {https://doi.org/10.1103/PhysRevD.68.103503}
  {\path{doi:10.1103/PhysRevD.68.103503}}.

\bibitem{NeferSenoguz:2008nn}
V.~N. Senoguz, Q.~Shafi, {Chaotic inflation, radiative corrections and
  precision cosmology}, Phys. Lett. B668 (2008) 6--10.
\newblock \href {http://arxiv.org/abs/0806.2798} {\path{arXiv:0806.2798}},
  \href {https://doi.org/10.1016/j.physletb.2008.08.017}
  {\path{doi:10.1016/j.physletb.2008.08.017}}.

\bibitem{Gong:2015qha}
J.-O. Gong, S.~Pi, G.~Leung, {Probing reheating with primordial spectrum}, JCAP
  05 (2015) 027.
\newblock \href {http://arxiv.org/abs/1501.03604} {\path{arXiv:1501.03604}},
  \href {https://doi.org/10.1088/1475-7516/2015/05/027}
  {\path{doi:10.1088/1475-7516/2015/05/027}}.

\bibitem{Akrami:2020zxw}
Y.~Akrami, S.~Casas, S.~Deng, V.~Vardanyan, {Quintessential $\alpha$-attractor
  inflation: forecasts for Stage IV galaxy surveys} (10 2020).
\newblock \href {http://arxiv.org/abs/2010.15822} {\path{arXiv:2010.15822}}.

\bibitem{Lewis_2000}
A.~Lewis, A.~Challinor, A.~Lasenby, Efficient computation of cosmic microwave
  background anisotropies in closed friedmann‐robertson‐walker models, The
  Astrophysical Journal 538~(2) (2000) 473–476.

\bibitem{Mortonson:2010er}
M.~J. Mortonson, H.~V. Peiris, R.~Easther, {Bayesian Analysis of Inflation:
  Parameter Estimation for Single Field Models}, Phys. Rev. D83 (2011) 043505.
\newblock \href {http://arxiv.org/abs/1007.4205} {\path{arXiv:1007.4205}},
  \href {https://doi.org/10.1103/PhysRevD.83.043505}
  {\path{doi:10.1103/PhysRevD.83.043505}}.

\bibitem{weinberg2008cosmology}
S.~Weinberg, Cosmology, Cosmology, OUP Oxford, 2008.

\bibitem{Lewis_2002}
A.~Lewis, S.~Bridle, Cosmological parameters from cmb and other data: A monte
  carlo approach, Physical Review D 66~(10) (Nov 2002).

\bibitem{Aghanim:2015xee}
N.~Aghanim, M.~Arnaud, M.~Ashdown, J.~Aumont, C.~Baccigalupi, A.~J. Banday,
  R.~B. Barreiro, J.~G. Bartlett, N.~Bartolo, et~al., Planck 2015 results. xi.
  cmb power spectra, likelihoods, and robustness of parameters, Astronomy \&
  Astrophysics 594 (2016) A11.

\bibitem{Kallosh_2015}
R.~Kallosh, A.~Linde, Planck, lhc, and $\alpha$-attractors, Physical Review D
  91~(8) (Apr 2015).

\bibitem{Kallosh_2014}
R.~Kallosh, A.~Linde, D.~Roest, Universal attractor for inflation at strong
  coupling, Physical Review Letters 112~(1) (Jan 2014).

\bibitem{Ferrara:2016fwe}
S.~Ferrara, R.~Kallosh, {Seven-disk manifold, $\alpha$-attractors, and $B$
  modes}, Phys. Rev. D 94~(12) (2016) 126015.
\newblock \href {http://arxiv.org/abs/1610.04163} {\path{arXiv:1610.04163}},
  \href {https://doi.org/10.1103/PhysRevD.94.126015}
  {\path{doi:10.1103/PhysRevD.94.126015}}.

\bibitem{Kallosh:2017wnt}
R.~Kallosh, A.~Linde, D.~Roest, Y.~Yamada, {$ \overline{D3} $ induced geometric
  inflation}, JHEP 07 (2017) 057.
\newblock \href {http://arxiv.org/abs/1705.09247} {\path{arXiv:1705.09247}},
  \href {https://doi.org/10.1007/JHEP07(2017)057}
  {\path{doi:10.1007/JHEP07(2017)057}}.

\bibitem{Kallosh:2017ced}
R.~Kallosh, A.~Linde, T.~Wrase, Y.~Yamada, {Maximal Supersymmetry and B-Mode
  Targets}, JHEP 04 (2017) 144.
\newblock \href {http://arxiv.org/abs/1704.04829} {\path{arXiv:1704.04829}},
  \href {https://doi.org/10.1007/JHEP04(2017)144}
  {\path{doi:10.1007/JHEP04(2017)144}}.

\bibitem{Renzi:2019ewp}
F.~Renzi, M.~Shokri, A.~Melchiorri, {What is the amplitude of the gravitational
  waves background expected in the Starobinski model?}, Phys. Dark Univ. 27
  (2020) 100450.
\newblock \href {http://arxiv.org/abs/1909.08014} {\path{arXiv:1909.08014}},
  \href {https://doi.org/10.1016/j.dark.2019.100450}
  {\path{doi:10.1016/j.dark.2019.100450}}.

\bibitem{SantosdaCosta:2020ont}
S.~Santos~da Costa, M.~Benetti, J.~Alcaniz, R.~Silva, R.~Neves, {Robustness of
  the Starobinsky inflationary model} (7 2020).
\newblock \href {http://arxiv.org/abs/2007.09211} {\path{arXiv:2007.09211}}.

\end{thebibliography}





\end{document}